\documentclass[prl,12pt]{revtex4}

\usepackage{amsmath}
\usepackage{amssymb}

\usepackage{graphicx}

\begin{document}

\title{Magnetic tunnel junctions with impurities}

\author{F. Kanjouri$^{1,2} $, N. Ryzhanova$^{1,3}$, B.  Dieny$^{3}$, N. Strelkov$^{1,3}$, A. Vedyayev$^{1,3}$}
\address{
$^1$Department of Physics, Moscow Lomonosov University, Moscow 119899, Russia\\
$^2$ Department of Physics, Yazd University, Yazd, Iran\\
$^3$CEA/Grenoble, Department de Recherche Fondamentale sur la Matiere Condensee, Spintec. 38054 Grenoble, France}

\begin{abstract}
\textbf{Abstract:} The influence of impurities, embedded into the
isolating spacer ($\mathbb{I}$) between two ferromagnetic
electrodes ($\mathbb{F}$), on the I-V curve and tunnel
magnetoresistance (TMR), is theoretically investigated. It is
shown, that the current and TMR are strongly enhanced in the
vicinity of the impurity under the condition that the energy of
the electron's bound state on the impurity is close to the Fermi
energy. If the position of the impurity inside the barrier is
asymmetric, e.g. closer to the one of the interfaces ${\mathbb F
}$/${\mathbb I }$ the I-V curve exhibits quasidiode behavior.
\end{abstract}

\maketitle

The magnetic tunnel junction (TMJ), consisting of two metallic
ferromagnetic electrodes separated by insulating barrier and
exhibiting tunnel magnetoresistance (TMR) of order 50\% attracts a
lot of attention~\cite{Moodera-Kinder_1,Parkin,Moodera-Mathon_1}
especially due to their possible application in MRAM (Magnetic
Random Access Memory). In the pioneer paper~\cite{Slonczewski} the
theory of TMR for the ideal (without defect) TMJ was developed.
Later it was shown~\cite{Tsymbal-Pettifor_1,Vedy-Bagrets_1} that
in presence of different types of defects within the barrier the
I-V current and TMR change dramatically. In these papers the
averaged over cross section of the system current was calculated.
However it is interesting to investigate the local (in vicinity of
the impurity) current and TMR, especially taking into account that
using the STM technique it is possible and was
realized~\cite{DaCosta-Herry} to span tunnelling current over the
cross section of TMJ. Recently the theory of local impurity
assisted tunnelling in TMJ was
developed~\cite{Tsymbal-Pettifor_2}. As a model of TMJ was taken
tight binding model and Kubo formalism was used for calculation of
spin-dependent tunnel current. In this paper the I-V curve was not
investigated in detail and besides that the dependence of
spin-dependent current on the position of cross section plane
relative to the position of impurity was not investigated.

In the presented paper we investigate the local distribution of
spin-dependent current for different positions of the cross
section plane and local I-V  curves of TMJ with single impurity
and for random distribution of impurities inside the barrier. We
adopted the free electron model with exchange splitting for
ferromagnetic electrodes and used nonequilibrium Keldysh
technique~\cite{Keldysh_eng} for calculating of nonlinear on
applied voltage transport properties.

We considered the model of TMJ as a three layers system, consisted
from two thick ferromagnetic electrodes ${\mathbb F }$ separated
by the insulating layer, ${\mathbb I }$ . Inside the barrier the
single nonmagnetic impurity with attracting potential was situated
at some distance from ${\mathbb F }$/${\mathbb I }$  interface.
The two cases were investigated : parallel and antiparallel
orientations of ${\mathbb F }$-layers magnetization.

The ${\mathbb F }$-electrodes are connected to the reservoirs with
chemical potentials $\mu_{1}$ and $\mu_{2}$ so that
$\mu_{2}-\mu_{1}=eV$, where $V$ is the applied voltage.

To calculate the current through the system we have to found
Keldysh  Green function $G^{-+}$ and advanced and retarded Green
functions $G^{A}$ and $G^{R}$. Solving the Dyson equation we found
that
\begin{multline}
G^{-+}(\textbf{r},\textbf{r}')=G_{0}^{-+}
(\textbf{r},\textbf{r}')+
\frac{G_{0}^{R}(\textbf{r},\textbf{r}_{0})
WG_{0}^{-+}(\textbf{r}_{0},\textbf{r}')}
{1-WG_{0}^{R}(\textbf{r}_{0},\textbf{r}_{0})}
+\frac{G_{0}^{-+}(\textbf{r},\textbf{r}_{0})
WG_{0}^{A}(\textbf{r}_{0},\textbf{r}')}
{1-WG_{0}^{A}(\textbf{r}_{0},\textbf{r}_{0})}
\\+\frac{G_{0}^{R}(\textbf{r},\textbf{r}_{0})
WG_{0}^{-+}(\textbf{r}_{0},\textbf{r}_{0})
WG_{0}^{A}(\textbf{r}_{0},\textbf{r}')}
{\left(1-WG_{0}^{R}(\textbf{r}_{0},\textbf{r}_{0})\right)
\left(1-WG_{0}^{A}(\textbf{r}_{0},\textbf{r}_{0})\right)}\label{Dyson}
\end{multline}
where $G_{0}^{-+} (\textbf{r},\textbf{r}'),G_{0}^{A}
(\textbf{r},\textbf{r}')$ and $G_{0}^{R} (\textbf{r},\textbf{r}')$
are the Green's functions for the system in the absence of the
impurity and the potential of the impurity $V$ was represented as
$\delta$-function:
$V(\textbf{r})=Wa_{0}^3\delta(z-z_{0})\delta$(\mbox{\boldmath
$\rho$}-\mbox{\boldmath $\rho_{0}$}),
$\textbf{r}_{0}=(\mbox{\boldmath $\rho_{0}$},z_{0})$ is the
position of the impurity, $a_{0}$ is it's effective radius, $W$ is
it's intensity. The explicit expressions for $G^A, G^R, G^{-+}$
have the following form:

\begin{eqnarray}
\begin{array}{l}
G_{0}^{R}(\textbf{r},\textbf{r}')=\int
d^2\kappa\frac{(-1)\mathrm{e}^{-i\mbox{\boldmath$\kappa$}(\mbox{\boldmath$\rho$}-\mbox{\boldmath$\rho$}')}}
{2\sqrt{q(z)q(z')}den}\left\{
  E(z_{2},z)\left[q(z_{2})+ik_{2}\right]+E^{-1}(z_{2},z)\left[q(z_{2})-ik_{2}\right]
  \right\}\\ \times\left\{
  E(z',z_{1})\left[q(z_{1})+ik_{1}\right]+E^{-1}(z',z_{1})\left[q(z_{1})-ik_{1}\right]
  \right\},\label{G0R}
 \end{array}
\end{eqnarray}

\begin{eqnarray}
\begin{array}{l}
G_{0}^{A}(\textbf{r},\textbf{r}')=\int d^2\kappa\frac{(-1)
  \mathrm{e}^{i\mbox{\boldmath$\kappa$}(\mbox{\boldmath$\rho$}-\mbox{\boldmath$\rho$}')}}{2\sqrt{q(z)q(z')}den^*}\left\{
  E(z_{2},z)\left[q(z_{2})-ik_{2}\right]+E^{-1}(z_{2},z)\left[q(z_{2})+ik_{2}\right]
  \right\}\\ \times\left\{
  E(z',z_{1})\left[q(z_{1})-ik_{1}\right]+E^{-1}(z',z_{1})\left[q(z_{1})+ik_{1}\right]
  \right\},\label{G0A}
 \end{array}
\end{eqnarray}

\begin{eqnarray}
\begin{array}{l}
G_{0}^{-+}(\textbf{r},\textbf{r}')=\int
d^2\kappa\frac{i4k_{1}q(z_{1})n_{L}\mathrm{e}^{-i\mbox{\boldmath$\kappa$}(\mbox{\boldmath$\rho$}-\mbox{\boldmath$\rho$}')}}
{\sqrt{q(z)q(z')}|den|^2}\left\{
  E(z',z_{2})\left[q(z_{2})+ik_{2}\right]+E^{-1}(z',z_{2})\left[q(z_{2})-ik_{2}\right]
  \right\}\\ \times\left\{
  E(z,z_{2})\left[q(z_{2})-ik_{2}\right]+E^{-1}(z,z_{2})\left[q(z_{2})+ik_{2}\right]
  \right\}\\
  \\+\int
d^2\kappa\frac{i4k_{2}q(z_{2})n_{R}\mathrm{e}^{-i\mbox{\boldmath$\kappa$}
(\mbox{\boldmath$\rho$}-\mbox{\boldmath$\rho$}')}}{\sqrt{q(z)q(z')}|den|^2}\left\{
  E(z_{1},z')\left[q(z_{1})+ik_{1}\right]+E^{-1}(z_{1},z')\left[q(z_{1})-ik_{1}\right]
  \right\}\\ \times\left\{
  E(z_{1},z)\left[q(z_{1})-ik_{1}\right]+E^{-1}(z_{1},z)\left[q(z_{1})+ik_{1}\right]
  \right\},\label{G0-+}
 \end{array}
\end{eqnarray}

where
\begin{eqnarray*}
\begin{array}{l}
q(z)=\sqrt{q_{0}^{2}+\kappa^{2}-\frac{2m}{\hbar^2}\frac{(z-z_{1})}{(z_{2}-z_{1})}eV},
\\\\k_{1}=\sqrt{\frac{2m}{\hbar^2}(\varepsilon-\Delta_{1})-\kappa^{2}},
\\\\k_{2}=\sqrt{\frac{2m}{\hbar^2}(\varepsilon-\Delta_{2}+eV)-\kappa^{2}},
\\\\den=\left\{
{E(z_{1},z_{2})\left[q(z_{2})-ik_{2}\right]\left[q(z_{1})-ik_{1}\right]
-E^{-1}(z_{1},z_{2})\left[q(z_{2})+ik_{2}\right]\left[q(z_{1})+ik_{1}\right]}\right\},
\\ \\E(z_{1},z_{2})\equiv \mathrm{e}^{\int_{z_{1}}^{z_{2}}
q(\tau)d\tau},
\end{array}
\end{eqnarray*}
\mbox{\boldmath$\kappa$} is the electron momentum perpendicular to
the plane of structure, $\varepsilon$ is the energy, $z_{1}$ and
$z_{2}$ are the positions of ${\mathbb F }$/${\mathbb I }$
interfaces, $\Delta_{1}$ and $\Delta_{1}$ denote   the positions
of energy band bottom  for spin up and down subbands.

\noindent  $n_L=f^0(\varepsilon)$ and $n_R=f^0(\varepsilon+eV)$
are Fermi distribution functions in the left and right reservoirs
and $\frac{\hbar^2q_{0}^2}{2m}$ height of potential barrier above
Fermi's level.

\begin{figure}[ht]
\includegraphics*[width=0.57\textwidth]{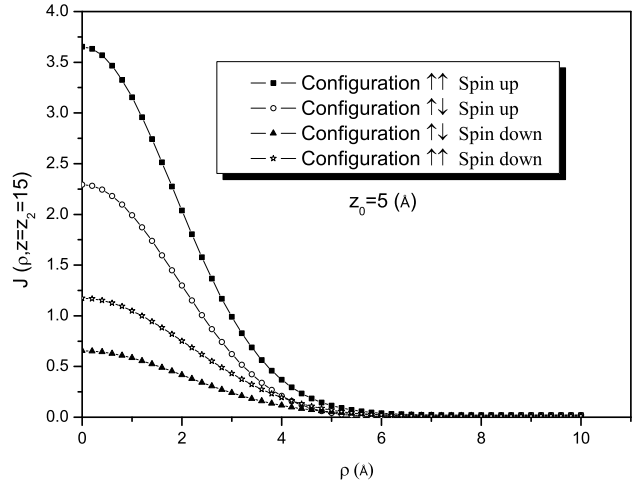}
\caption{Dependence of the current for different spin channels and
P and AP configuration on the distance from the impurity in the
plane of the structure at z=15 \AA. $k_{F}^\uparrow=1.1
~{\mbox\AA}^{-1}$, $k_{F}^\downarrow=0.6 ~{\mbox\AA}^{-1}$,
$q_{0}=1.0 ~{\mbox\AA}^{-1}$ }. \label{fig:fig1}
\end{figure}
\begin{figure}[h]
\includegraphics*[width=0.57\textwidth]{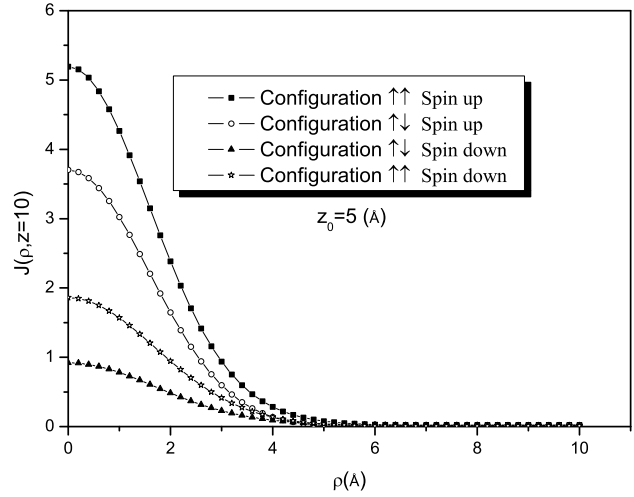}
\caption{The same dependence at $z=10$ \AA .} \label{fig:fig2}
\end{figure}

\begin{figure}[h]
\includegraphics*[width=0.57\textwidth]{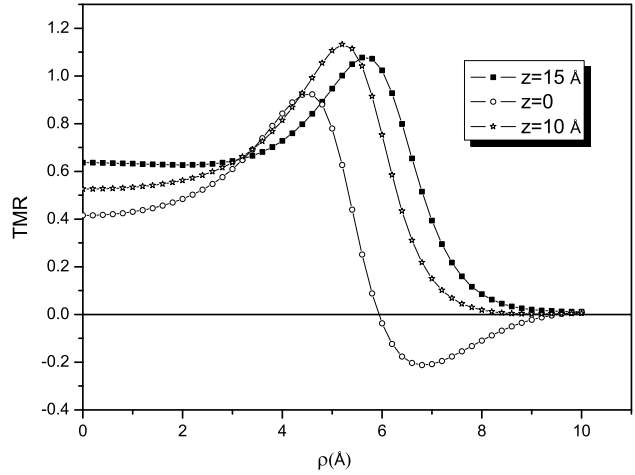}
\caption{Dependence of TMR on the distance from the impurity in
the plane of the structure at different $z$. For parameters see
Fig.\ref{fig:fig1}} \label{fig:fig3}
\end{figure}

\begin{figure}[h]
\includegraphics*[width=0.5\textwidth,angle=-90]{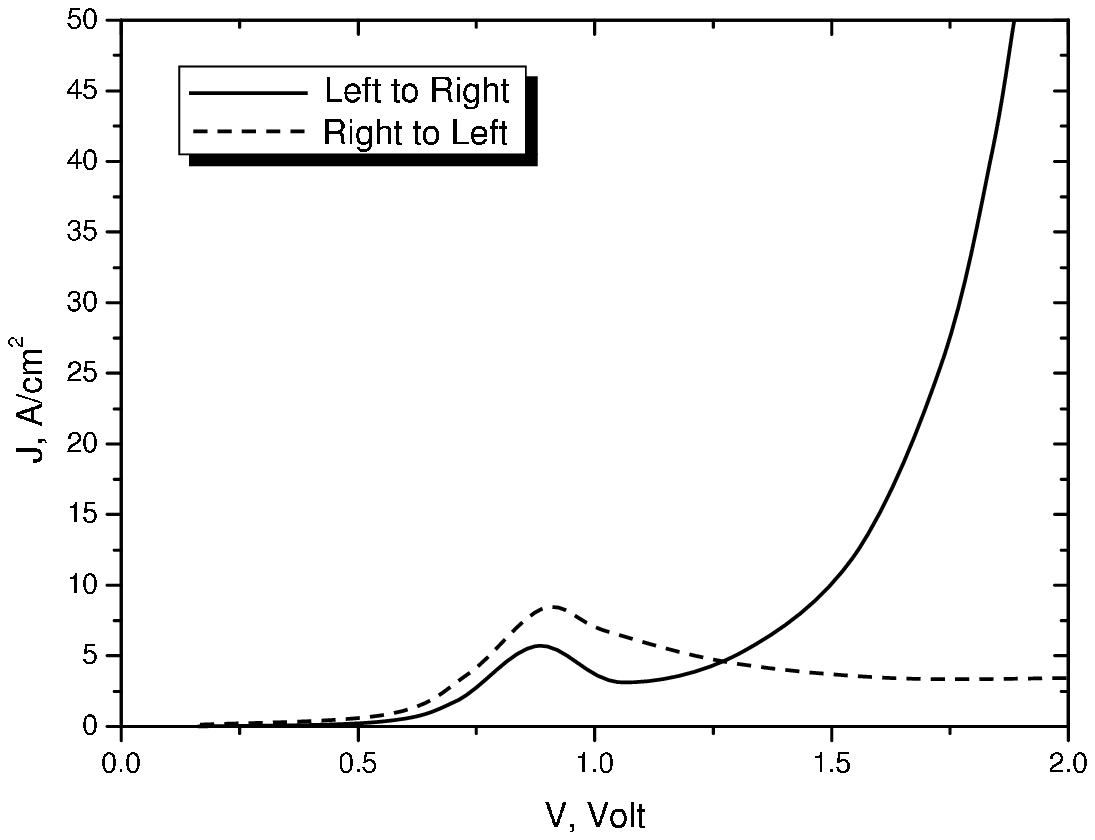}
\caption{ Local I-V curve at $\rho=\rho_{0}$ and $z=15$ \AA ~for
the case of single impurity.} \label{fig:fig4}
\end{figure}
\begin{figure}[ht]
\includegraphics*[width=0.5\textwidth,angle=-90]{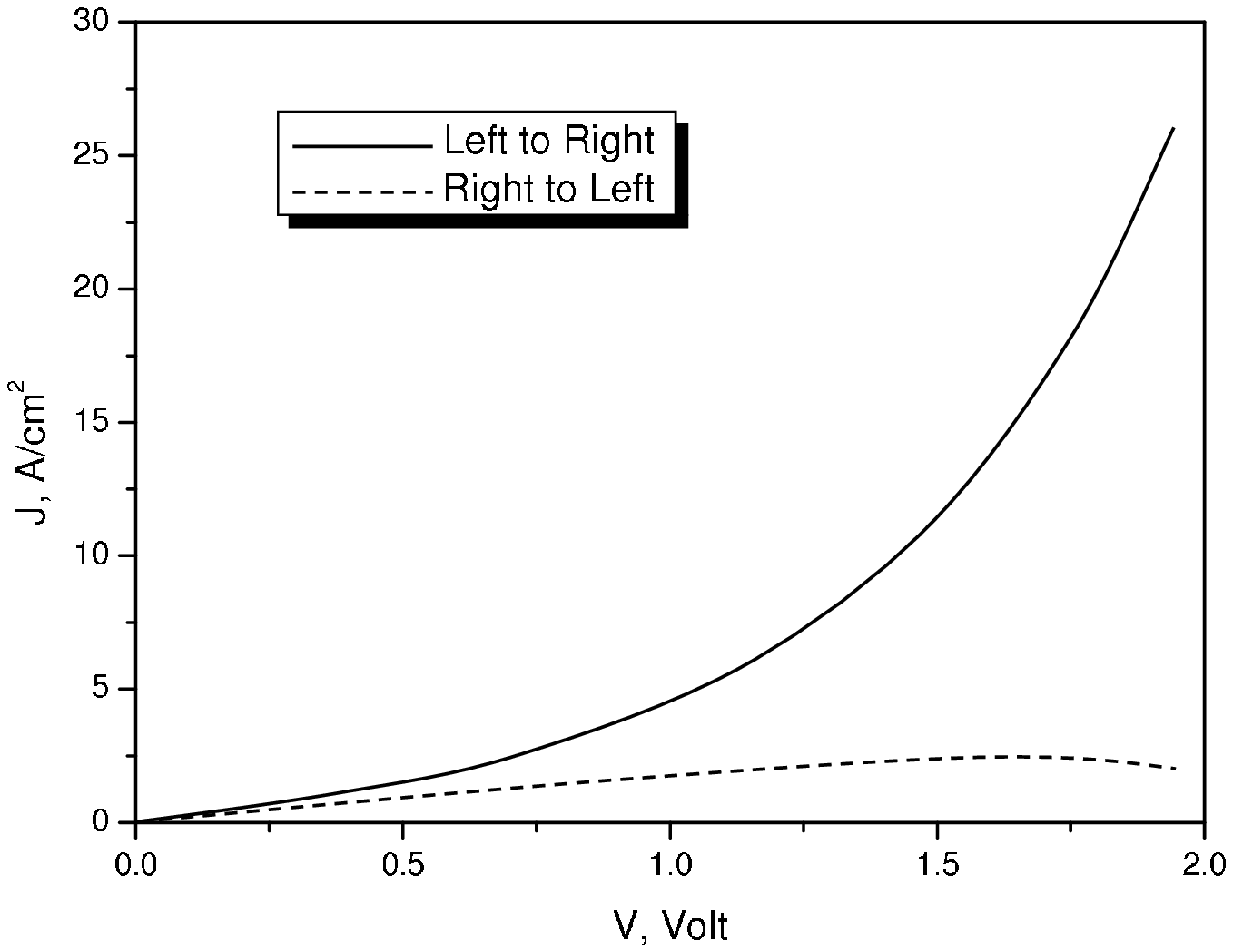}
\caption{I-V curve in the case of the layer of impurities at
$z_{0}=3$\AA ~and $x=0.5$.} \label{fig:fig5}
\end{figure}

In (\ref{Dyson}),(\ref{G0R}),(\ref{G0A}) and (\ref{G0-+})
\mbox{\boldmath$\rho$} and $z$ are in the plane and perpendicular
to the plane coordinates, and we consider that $z$ and $z_{0}$ are
situated within the barrier. We have to take into account that all
Green functions are matrixes in spin space. We have consider
$k_{1F}^\uparrow$, $k_{1F}^\downarrow$, $k_{2F}^\uparrow$,
$k_{2F}^\downarrow$ are Fermi wave vectors of electron with
spin$\uparrow(\downarrow)$ in the left and right
$\mathbb{F}$-electrodes. The current was calculated, using the
following expression:

\begin{equation}
j_{z}(\rho,z)=\frac{e\hbar}{2m}\int
d\varepsilon\left(\frac{\partial G^{-+}(z,\rho;z',\rho)}{\partial
z'}-\frac{\partial G^{-+}(z,\rho;z',\rho)}{\partial
z}\right)_{z=z'}\label{Jz}
\end{equation}

On the Fig.\ref {fig:fig1} and Fig.\ref {fig:fig2}  the
dependencies of the currents in different channels ( up and down
spin) on coordinate $\rho-\rho_{0}$ at one interface
$\mathbb{I}$/$\mathbb{F}$ ($z_{2}=15$\AA) (another interface is at
$z_{1}=0$) and inside the barrier at $z=10$ are shown. Position of
the impurity is at $\rho_{0}=0$ and $z_{0}=5$\AA.

So in the vicinity of the impurity the hot spot of the radius
approximately equal 6\AA ~may be observed and the value of the
current density in the center of the hot spot exceeds the value of
the background current on several orders of magnitude. On the
Fig.\ref {fig:fig3}  the TMR dependence on the distance from the
impurity at different $z$ is shown. It is interesting that the
value of TMR in vicinity of the impurity exceeds it's background
value (TMR for the ideal structure is equal $0.013$) more over
then order of magnitude and  for some cases it exists the region
of $\rho-\rho_{0}$ where TMR becomes negative.

Now on Fig.\ref {fig:fig4}  the I-V current for positive and
negative applied voltage is shown. This curves are quite
asymmetric on the sign of the voltage. It is connected with
asymmetry of position of the impurity inside of the barrier. We
choose the potential of impurity so that bound (resonance) state
of electrons with spin up located near Fermi level for the
positive applied voltage $=1.2 ~V$, and for negative voltage the
position of this bound state lies below Fermi level. This diode
behavior is demonstrated for the single impurity and to have the
possibility to use this diode effect in practice we have to
investigate the case of the finite concentration of impurities.

In this case we consider the same magnetic tunnel barrier
structure with monolayer of impurities of finite atomic
concentration $x$, situated closer to the one of ${\mathbb F
}$/${\mathbb I }$ interface. To solve the problem as a first step
we have to find coherent potential and effective Keldysh Green
function $G_{\mathrm{eff}}^{-+}$. Solving the Dyson equation in
the Keldysh space we got the following expression for
$G_{\uparrow\uparrow}^{-+AP}$:
\begin{multline}
G^{-+}\left({z,z'}\right)=G_0^{-+}\left({z,z'}\right)+
\frac{{G_0^{-+}\left({z,z_0 }\right)\Sigma^A G_0^A \left({z_0 ,z'}\right)}}
{{1-G_0^A\left({z_0 ,z_0 }\right)\Sigma^A}}
+\frac{{G_0^R \left({z,z_0}\right)\Sigma^R G_0^{-+} \left({z_0 ,z'}\right)}}
{{1-G_0^R \left( {z_0 ,z_0 } \right)\Sigma^R }}-\\
\frac{{G_0^R \left( {z,z_0 } \right)\Sigma^{-+} G_0^A \left({z_0
,z'}\right)}} {{\left({1-G_0^A\left({z_0
,z_0}\right)\Sigma^A}\right)\left({1-G_0^R\left({z_0,z_0}\right)\Sigma^R}\right)}}+
\frac{{G_0^R \left({z,z_0}\right)\Sigma^R
G_0^{-+}\left({z_0,z_0}\right)\Sigma^A G_0^A
\left({z_0,z'}\right)}}
{{\left({1-G_0^A\left({z_0,z_0}\right)\Sigma^A}\right)\left({1-G_0^R\left({z_0,z_0}\right)\Sigma^R}\right)}}\label{G-+}
\end{multline}
where $\Sigma^{R(A)}$ are the coherent potential (C.P.) for the
retarded and advanced Green functions, which has to be found from
the C.P.A equation:
\begin{equation}
\bar t=(1-x)
\frac{(\varepsilon^A-\Sigma)}{1-(\varepsilon^A-\Sigma)G_{\mathrm{eff}}(z_{0},\rho_{0};z_{0},\rho_{0})}
+(x)\frac{(\varepsilon^B-\Sigma)}{1-(\varepsilon^B-\Sigma)G_{\mathrm{eff}}(z_{0},\rho_{0};z_{0},\rho_{0})}
=0\label{t}
\end{equation}
where $\varepsilon^A$ and $\varepsilon^B$ are the onsite energies
of the host ($Al_2O_3$) and the impurity ($Al$) and
$\Sigma^{-+}=\frac{i}{2}(n_R+n_L)(\Sigma^R-\Sigma^A)$.

Now to calculate I-V curve we may use the found $G_{\alpha
\alpha}^{-+P(AP)}$, substituting it into the
expression~(\ref{Jz}).

On the Fig.\ref{fig:fig5} the I-V curve for AP configuration is
shown, and the asymmetry of the curve on the sign of applied
voltage is clearly seen.

Such a structure may be prepared if to sputter thin layer of $Al$
on the $\mathbb{F}$-electrode, then oxidise it, after sputter
thicker layer of $Al$ and oxidise it from the top side not
completely. So some thin layer of the random alloy
$Al_xAl_2O_{3(1-x)}$ is situated inside the more or less ideal
insulator $Al_2O_3$ at the distance close to the first
$\mathbb{F}$/$\mathbb{I}$ interface.

The work was partly supported by the Russian fund of fundamental
research (grant N 04-02-16688a).

\bibliography{impur}

\end{document}